\documentclass[a4paper,12pt]{article}

\usepackage{palatino}
\usepackage{epsfig}
\usepackage{amsmath}
\usepackage[footnotesize]{caption}
\usepackage{fancyhdr}
\usepackage{subfigure}
\usepackage{booktabs}

\title{Beam Excited Signals in a Cavity BPM}
\author{
A.~Lyapin,
{\em University College London, UK}\\
}

\date{}

\graphicspath{{./plots/}}

\begin{document}
\thispagestyle{empty}
\maketitle
\begin{abstract}
A beam traversing a cavity BPM excites many modes, of which we are usually interested in the first lower modes with the strongest excitation, which are usually below the beampipe cut-off frequency. Besides the bunch charge, their signals may depend on the bunch offset, slope and tilt. Due to de-phasing between the bunch and the excited field, the excitation also has a non-linear dependence on the width of the cavity gap and the bunch length. In this note I give estimates for each component of the fundamental (the fist monopole, or common) mode, and the first dipole mode using a two-bunch approximation.
\end{abstract}

\parskip=2mm
\pagestyle{fancyplain}
\renewcommand{\sectionmark}[1]{\markright{\thesection\ #1}}
\lhead[\fancyplain{}{\bfseries\thepage}]{\fancyplain{}{\bfseries\rightmark}}
\rhead[\fancyplain{}{\bfseries\leftmark}]{\fancyplain{}{\bfseries\thepage}}
\cfoot{}
\pagenumbering{arabic}


A bunch of charged particles crossing a cavity BPM excites an infinite number of cavity modes. For some modes the excitation is very weak due to their field configurations. Modes with the resonant frequencies higher than the beampipe cut-off escape from the cavity through the beampipe. But the modes that are used for beam position and phase measurements are designed to be below cut-off.

Even a relativistic bunch of particles (and here I assume $v=c$) takes some timeto cross the gap of the cavity, and, since the gap width is usually comparable with the wavelength of the mode in question, a de-phasing between the excited field and the bunch starts occuring. This leads to transient effects including non-linear dependence between the gap width and the signal amplitude, and bunch angle and tilt sensitivity. A similar effect takes place when the length of the bunch is comparable with the wavelength.

I use a few more approximations in this note. Firstly, I assume that the bunch has a Gaussian shape. In that case it can be represented by two point charges carrying a half of the total bunch charge each and separated by the RMS bunch length $\sigma$. Secondly, I assume that the excitation of the dipole mode is linear vs. beam offset. At the first glance, it seems that this approximation is only valid for small offsets, but the presence of the beampipe modifies the fields so that the linearity is significantly improved. The last approximation is that the fields only exist within the caivty gap. This assumption does affect the result, but since the field quickly decays in the beampipe for the trapped modes, the error is not likely to exceed 10-20\%. Some techniques allow for matching the fields of the cavity mode and the evanescent waveguide mode, but require more elaborate maths.

In order to compare the excitation, let us calculate the voltage corresponding to each component excited in the cavity. We integrate the electric field among the trajectory:

\begin{equation}
\label{gen_voltage}
V=\int\limits_{-\infty}^{+\infty}E\mbox{d}z=\int\limits_{-L/2}^{+L/2}E_0 e^{jkz}\mbox{d}z,
\end{equation}

\noindent where $L$ is the cavity length (or the width of the cavity gap), $k$ is the wave number for the mode and $E_m$ is the amplitude of the electric field.
 Rewriting this equation for our two-bunch model, we get:

\begin{equation}
V_0=\frac{qE_0}{2}\int\limits_{-L/2}^{+L/2} \left[e^{jk(z-\frac{\sigma}{2})} + e^{jk(z+\frac{\sigma}{2})}\right]\mbox{d}z.
\end{equation}

\noindent Integrating and rearranging this equation using Euler's formula, we obtain the voltage for the monopole mode:

\begin{equation}
V_0=qE_0L \cos\Big(\frac{k\sigma}{2}\Big) \frac{\sin kL/2}{kL/2}.
\end{equation}

\noindent The term $\frac{\sin kL/2}{kL/2}$ is known as the transit time factor and takes into account de-phasing between the field and the bunch while it crosses the gap. The cosine term takes care of the similar effect occuring due to a finite bunch length.

For the dipole mode the field depends on the offset $x$: $E = E_1xe^{jkz}$, hence:

\begin{equation}
\label{v_dipole}
V_1=E_1\int\limits_{-L/2}^{+L/2}xe^{jkz}\mbox{d}z.
\end{equation}

In case the offset is constant (the trajectory is parallel to the cavity axis) the voltage is simply:

\begin{equation}
\label{v_x}
V_x=qE_1x\int\limits_{-L/2}^{+L/2}\left[e^{jk(z-\frac{\sigma}{2})} + e^{jk(z+\frac{\sigma}{2})}\right]\mbox{d}z=qE_1xL \cos\Big(\frac{k\sigma}{2}\Big) \frac{\sin kL/2}{kL/2}.
\end{equation}

Now let's assume we have a bunch traveling through the cavity center but with a small slope $\theta$ to the axis, so that the offset is changing as $x=z\theta$. Equation (\ref{v_dipole}) is then modified to

\begin{equation}
\label{v_theta}
V_{\theta}=qE_1\theta\int\limits_{-L/2}^{+L/2}z \left[e^{jk(z-\frac{\sigma}{2})} + e^{jk(z+\frac{\sigma}{2})}\right] \mbox{d}z = j\frac{qE_1\theta L}{k} \cos\Big(\frac{k\sigma}{2}\Big) \frac{\sin kL/2}{kL/2} \left[ 1 - \frac{kL}{2} \cot \frac{kL}{2} \right].
\end{equation}

\noindent The imaginary unit $j$ in this equation signals that the inclination caused component has a 90$^0$ phase offset with respect to the displasement components.

In order to estimate the dipole mode response to a tilted bunch, we assume that the tilt $\alpha$ is small, so that the offset of the charges is $x=\pm\sigma\alpha/2$. This way (\ref{v_dipole}) becomes

\begin{equation}
V_{\alpha}=\frac{qE_1}{2} \frac{\alpha\sigma}{2} \int\limits_{-L/2}^{+L/2}\left[ e^{jk(z-\frac{\sigma}{2})} + e^{jk(z+\frac{\sigma}{2})} \right]\mbox{d}z = -jqE_1L \frac{\alpha\sigma}{2} \sin\Big(\frac{k\sigma}{2}\Big)  \frac{\sin kL/2}{kL/2}.
\end{equation}

Now, let us introduce the senstivity defined as the output signal generated by ashort bunch with a charge of 1~nC and, for the dipole mode, an offset of 1~mm. The sensitivity depends on the coupling strength and the shunt impedance of the cavity, so for the monopole mode we have

\begin{equation}
S_0 = \frac{\omega_0}{2} \sqrt{ \frac{Z}{Q_{ext}^{(0)}} \Big(\frac{R}{Q}\Big)_{x=0} }\cdot 10^{-9},
\end{equation}

\noindent in [V/nC] and for the dipole

\begin{equation}
S_1 = \frac{\omega_1}{2} \sqrt{ \frac{Z}{Q_{ext}^{(1)}} \Big(\frac{R}{Q}\Big)_{x=1 mm} }\cdot 10^{-9} \cdot 10^{-3}
\end{equation}

\noindent in [V/mm/nC].

Since the voltage coupled out from the cavity is proportional to the voltage excided in it, we can introduce our sensitivities into the above equations and get:

\begin{equation}
V_{out}^{(0)}=S_0 \cos\Big(\frac{k\sigma}{2}\Big) \hspace{1mm} q[nC];
\end{equation}

\begin{equation}
V_{out}^{(x)}=S_1 \cos\Big(\frac{k\sigma}{2}\Big) \hspace{1mm} q[nC] \hspace{1mm} x[mm];
\end{equation}

\begin{equation}
V_{out}^{(\theta)}= \frac{j}{k} S_1 \cos\Big(\frac{k\sigma}{2}\Big) \left[ 1 - \frac{kL}{2} \cot \frac{kL}{2} \right] \hspace{1mm} q[nC] \hspace{1mm} \theta[mrad];
\end{equation}

\begin{equation}
V_{out}^{(\alpha)}= -\frac{j\sigma}{2} S_1 \sin\Big(\frac{k\sigma}{2}\Big) \hspace{1mm} q[nC] \hspace{1mm} \alpha[mrad];
\end{equation}

These equations allow to quickly estimate the signals for a given cavity, or relative contributions of the dipole mode components for different gap lengths, frequencies and bunch lengths.

If you spot any mistakes or typos, please, report them to me at al@hep.ucl.ac.uk

\end{document}